\documentclass[conference]{IEEEtran}
\IEEEoverridecommandlockouts

\usepackage{amssymb,amsthm,amsmath,array}
\usepackage{graphicx}
\usepackage{xspace}
\usepackage[sort&compress, numbers]{natbib}
\usepackage{stmaryrd}
\usepackage{xcolor,soul,framed} \usepackage{mathtools}
\usepackage{float}
\usepackage{textcomp}

\usepackage{subfigure}
\usepackage{multirow}    
\usepackage{multicol}         
\usepackage{bbding}
\usepackage{enumitem}
\usepackage{algorithm}
\usepackage{algpseudocode}
\usepackage{booktabs}
\usepackage{tabularx}
\usepackage{makecell}
\usepackage [latin1]{inputenc}
\usepackage{bm}

\begin{document}
\title{Environment Reconstruction in Multi-Bounce Channels with Array Partial Blockage  \\
\thanks{This research was funded by the Luxembourg National Research Fund (FNR) through the BRIDGES project MASTERS under grant BRIDGES2020/IS/15407066 and the CORE INTER project SENCOM under C20/IS/14799710/SENCOM.}
}
\author{\IEEEauthorblockN{
        Yuan Liu\IEEEauthorrefmark{1}, 
        Linlong Wu\IEEEauthorrefmark{1},  
        Xuesong Cai\IEEEauthorrefmark{2}, and
        M.~R.~Bhavani Shankar\IEEEauthorrefmark{1}
    }
    \IEEEauthorblockA{
        \IEEEauthorrefmark{1} Interdisciplinary Centre for Security, Reliability and Trust (SnT), University of Luxembourg, L-1855, Luxembourg.\\
        \IEEEauthorrefmark{2} School of Electronics, Peking University, Beijing, 100871, China.
        }
}

\maketitle
\begin{abstract}
Extremely-large antenna arrays (ELAA) are important in applications requiring high angular resolution. 
However, a prominent issue is the spatial non-stationary (SNS) channels due to partial blockage to the ELAA. 
In this paper, we address the scatterer localization and subsequent environment reconstruction considering partially blocked SNS channels.  
Specifically, the SNS effects are parametrically modeled through spatial-varying amplitudes with sparsity. Based on the established signal model, the graph-based dictionary-aided multi-bounce space-alternating generalized expectation-maximization (GM-SAGE) algorithm is applied to estimate the channel parameters and the channel sparsity is empirically detected along with amplitude estimation.
To validate the proposed approach, we generate multi-bounce paths through ray tracing (RT) simulations, where the SNS channels caused by partial blockage could be configured flexibly.  The simulation results demonstrate the robustness of the proposed approach in dealing with the SNS channels. 
\end{abstract}

\section{Introduction}
Radio-based simultaneous localization and mapping (SLAM) finds various applications such as internet of things (IoTs), indoor robotics, and automotive driving \cite{10553198,10175555}.  
These applications rely on reconstructing the targeted environment by extracting the geometry information from the various propagation paths. 
Conventionally,  the literature relies on the direct (i.e. one-bounce) path from the scatterer to achieve this objective \cite{8823946,wang2021room}.  
However, the rich multi-bounce paths via multiple reflection/scattering, are inherent to the wireless channels and hard to ignore, especially in complex indoor environments\cite{ling2017experimental}.
Besides, using solely the one-bounce model can lead to incorrect scatterer estimations due to model mismatch, especially in cases where multi-bounce signals are relatively strong \cite{Yuan_TWC25}. 

In addition to the multi-bounce paths, 
extremely-large antenna array (ELAA) wideband systems are highly desired given their enhanced range and angle resolutions \cite{10445208}. However, due to the large aperture dimensions, the near-field effects, such as spherical wavefront or range-angle coupling,  arise in the modeling of ELAA \cite{7501567,10694490}.  Another important artifact of the ELAA systems is the spatial non-stationary (SNS) in the path gain across the array \cite{8713575}. 
%
Many reasons can be attributed to the SNS, including unresolved paths, partial blockage of the array, wave propagation mechanism, and imperfect hardware coupling \cite{9215972,8937499,8713575}. 

In this paper, we undertake scatterer localization to reconstruct radio maps using multi-bounce propagation. 
Due to the partial blockage of ELAA, conventional joint multi-parameter estimation algorithms, including the iterative framework of space-alternating generalized expectation-maximization (SAGE) \cite{fleury1999channel}, fail to perform the unknown SNS parameters. 
Different from the existing works, the proposed graph-based dictionary-aided multi-bounce SAGE (GM-SAGE) is a two-stage algorithm \cite{Yuan_TWC25}.
The channel parameters of the reference channel, e.g., delay and amplitude, are first estimated; then the joint multi-channel estimation is performed based on the delay of each multipath in the reference channel. 
Thus, the SNS amplitude in blocked channels can be captured.    


\section{Signal model and problem formulation}
\subsection{ Spherical Wavefront Parametric Channel Model}\label{Sec:signal_model}
Let us consider an ELAA system that operates a non-overlapping frequency division multiplexing (FDM) waveform, where $M$ and $N$ are the number of antennas utilized at the Tx and Rx sides, respectively. Furthermore, the FDM system utilizes $P$ sub-bands within a frame, with each sub-bandwidth being $f_s$.
%
Let $\mathbf{z} \in \mathbb{C}^{MNP \times 1}$ denote the baseband equivalent channel, with $m = 1, 2, ..., M$ and $n = 1, 2, ..., N$ denoting the antenna indices of the Tx and the Rx array respectively, and $p = 1, 2, ..., P$ denoting the index of the sub-band.
A unified parametric model for both near field and far field consists of $L$ multipath components (MPCs)  taking the form
\begin{align}\label{eq:ch_1}
    \mathbf{z} &= \sum_{l=1}^{L} \mathbf{z}_l \triangleq \sum_{l=1}^{L} \alpha_{l} \left[\left(\mathbf{\gamma}_{l} \odot  \Delta \mathbf{\alpha}_{l} \right)  \otimes \mathbf{1} \right]\odot \mathbf{a}_l(\mathbf{\tau})
\end{align} 
where $\odot, \otimes$ denote the Hadamard and Kronecker products respectively, $l = 1, 2, ..., L$ denotes the multipath index, $\alpha_{l}$ denotes the amplitude of the reference channel of the $l$th path, $\mathbf{\gamma}_{l}$ and $\mathbf{\alpha}_{l}$ represent the SNS characteristics. Here,  
$\mathbf{\gamma}_{l} \in \mathbb{C}^{MN \times 1} = [\gamma_{m,n,l}]_{MN \times 1}$ is the sparsity parameter vector indicating the physical blockage of the $l$th path, with  the channel between the $m$th Tx and the $n$th Rx, $\gamma_{m,n,l}$, being specified as 
\begin{equation} \label{eq:ch_1_3}
\gamma_{m,n,l} = 
\begin{cases}
0,   & \text{the $l$th path is blocked } \\
1,   & \text{the $l$th path is unblocked }.
\end{cases}  
\end{equation}
$\Delta \mathbf{\alpha}_{l} \in \mathbb{C}^{MN \times 1} = [\alpha_{m,n,l}]_{MN \times 1}$ denotes the SNS amplitude attenuation across the array, and $\mathbf{1} \in \mathbb{C}^{P \times 1}$ is a one vector.
$\mathbf{a}_l(\tau) \in \mathbb{C}^{ MNP \times 1} $ denotes the steering vector of the absolute delay under the spherical assumption\footnote{This model is a general expression of both near field and far field. When the distance is large enough, this steering vector will approximate to the planer wavefront \cite{Yuan_TWC25}.} as
\begin{equation} \label{eq:ch_vector_1}
        \mathbf{a}_l(\mathbf{\tau})  =  
    \left[ e^{\tau_{m,n,l}} \right]_{MN \times 1} \otimes \left[e^{-j2\pi f_p}\right]_{P \times 1},
\end{equation}
where $f_p = p f_s$ and $\tau_{m,n,l}$ denotes delay as
\begin{equation} \label{eq:delay_0}
\tau_{m,n,l} =  \frac{d_{m,n,l}}{c},
\end{equation}
where $c$ is the speed of light.  Further, $d_{m,n,l}$ is characterized by the coordinates of Tx and Rx array elements, and scatterers. Using graph-based multi-bounce model in \cite{Yuan_TWC25}, 
$d_{m,n,l}$ is calculated as
\begin{equation} \label{eq:delay_1_1}
\begin{aligned}
d_{m,n,l} = 
\begin{cases}
 \| \mathbf{r}_{l} - \mathbf{r}_{\text{Tx},m} \|  +  
 \| \mathbf{r}_{l} - \mathbf{r}_{\text{Rx},n} \|, ~~ \text{one-bounce path} \\
 \| \mathbf{r}_{l}^{1} - \mathbf{r}_{\text{Tx},m} \|  +  ... + \| \mathbf{r}_{l}^{K-1} - \mathbf{r}_{l}^{K} \|     \\   
   ~~~~~~~ + \| \mathbf{r}_{l}^{K} - \mathbf{r}_{\text{Rx},n} \|,  
 ~~~~~~~~ \text{multi-bounce path} 
\end{cases}  
\end{aligned}
\end{equation}
where for a one-bounce path, the signal is reflected off a single scatterer with coordinates $\mathbf{r}_{l}$. For a $K$-bounce path, $K \geq 2$, the signal propagates sequentially via $K$ scatterers, $\{ \mathbf{r}_{l}^1, ..., \mathbf{r}_{l}^k \}_{k=1}^{K}$ denotes coordinates of a set of $K$ scatterers.
%
\subsection{Problem formulation}
Given the measurement data, $\mathbf{y} \in \mathcal{C}^{MNP \times 1}$, our task is to jointly locate the coordinates of the scatterers and estimate the channel impacted by SNS. 
For this multipath channel estimation problem, we first define the $L-$ hidden multipaths in the measurement as
\begin{equation} 
\begin{aligned} \label{eq:prob_0}
    \mathbf{y} \triangleq \sum_{l=1}^{L} {\mathbf{z}_l}(\mathbf{\theta}_l) + \beta_l \mathbf{w},
\end{aligned}
\end{equation}
with $\mathbf{w}$ denotes the noise and $\sum_{l=1}^{L} \beta_l^2 =1$ to constrain the noise, and  
$\mathbf{\theta}_l$ denotes the channel parameters of the $l$th path.
Then, we formulate the problem as
\begin{align}  \label{eq:prob_1} 
    \mathcal{P}_1 ~~  & \arg \underset{ \mathbf{\theta}_l  }{\operatorname{ \min}}  \| \mathbf{y} - \sum_{l=1}^{L} {\mathbf{z}_l}_{|{  \mathbf{\theta}_l  }} \|,  
\end{align}
where $\mathbf{z}_l$ is the equivalent parametric model of the $l$th path defined in \eqref{eq:ch_1}, and  
\begin{equation} \label{eq:parameter}
\begin{aligned}
 \mathbf{\theta}_l = 
 \begin{cases}
 [\alpha_l, \Delta\mathbf{\alpha}_l, \mathbf{\gamma}_{l}, \mathbf{r}_{l}], & \text{one-bounce path} \\
 [\alpha_l, \Delta\mathbf{\alpha}_l, \mathbf{\gamma}_{l}, \{ \mathbf{r}_{l}^{1}, \mathbf{r}_{l}^{2} \} ], & \text{two-bounce path}     \\    [\alpha_l, \Delta\mathbf{\alpha}_l, \mathbf{\gamma}_{l}, \mathbf{\tau}_{l}] , & \text{high-bounce path}.
\end{cases}  
\end{aligned}
\end{equation}
For paths with more than three bounces, i.e., high-bounce paths, identifying the exact coordinates of scatterers would result in ambiguity \cite{Yuan_TWC25}.  
Hence, we only estimate the delay vector $\mathbf{\tau}_l$ for high-bounce paths. 

\section{GM-SAGE based scatterer localization}
\begin{figure*}[ht]
     \centering
          \subfigure[One-bounce reconstruction with blockage ]{\includegraphics[width=0.33\textwidth]{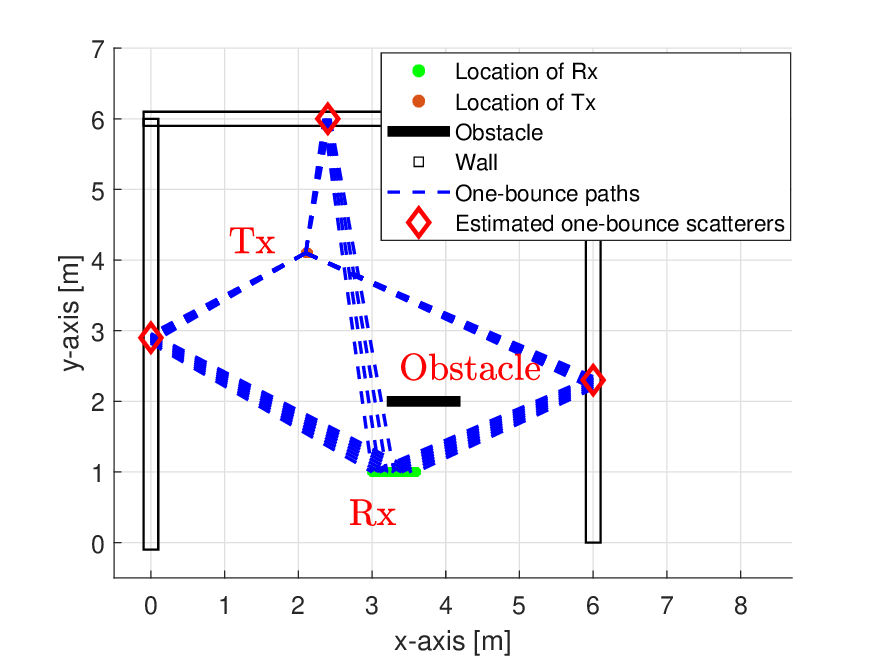}}
                        \hfill
    \subfigure[Two-bounce reconstruction with blockage]{\includegraphics[width=0.33\textwidth]{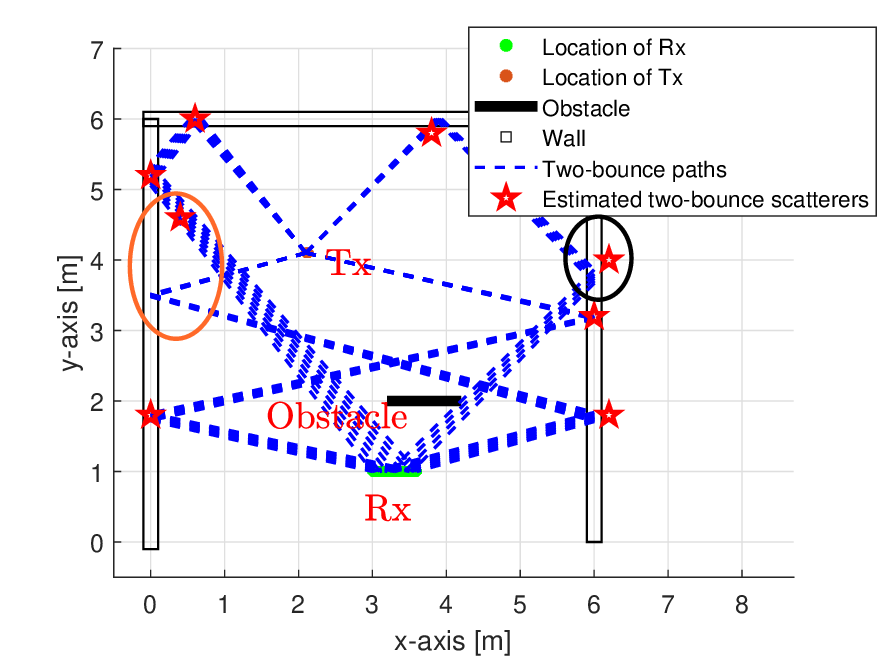}}
    \hfill
    \subfigure[Estimated SNS amplitude with blockage]{\includegraphics[width=0.32\textwidth]{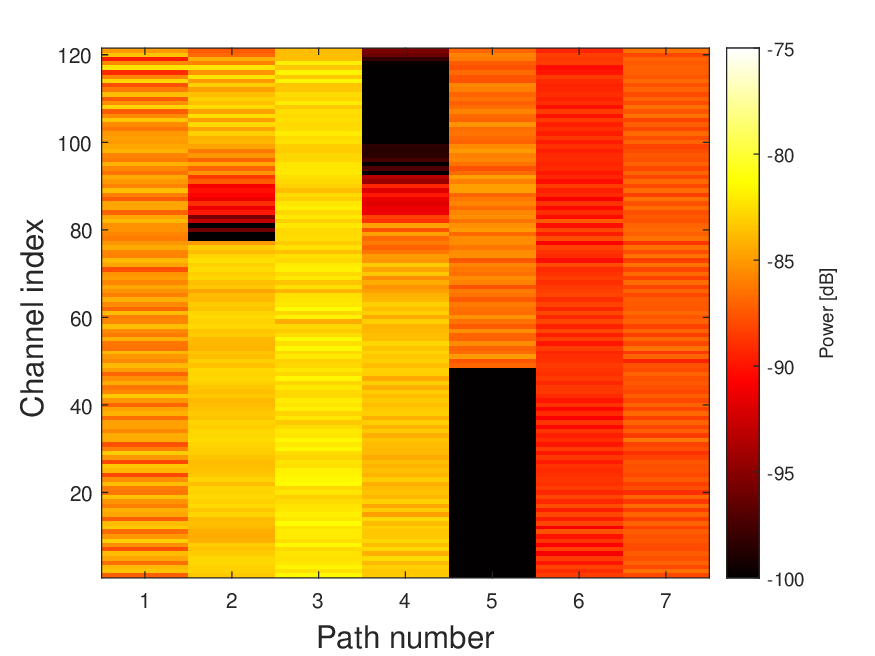}}
                        \hfill
        \subfigure[One-bounce reconstruction without blockage ]{\includegraphics[width=0.33\textwidth]{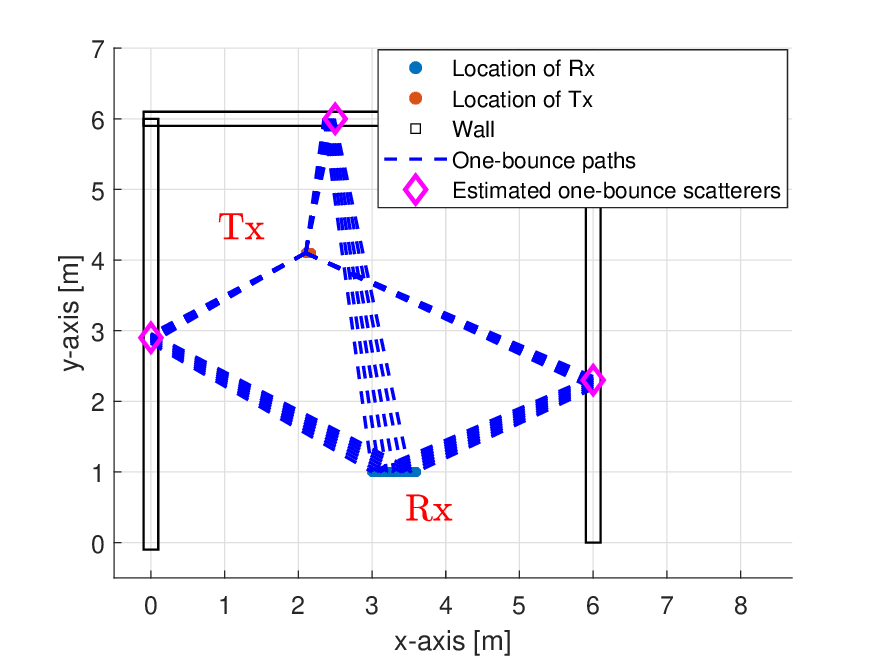}}
                        \hfill
        \subfigure[Two-bounce reconstruction without blockage]{\includegraphics[width=0.33\textwidth]{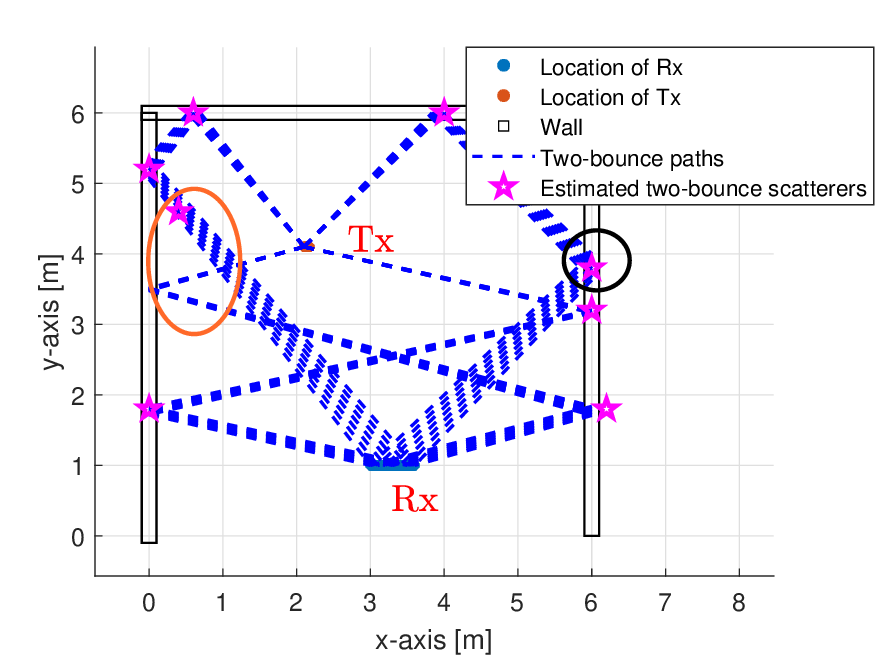}}
    \hfill
    \subfigure[Estimated SNS amplitude without blockage  ]{\includegraphics[width=0.32\textwidth]{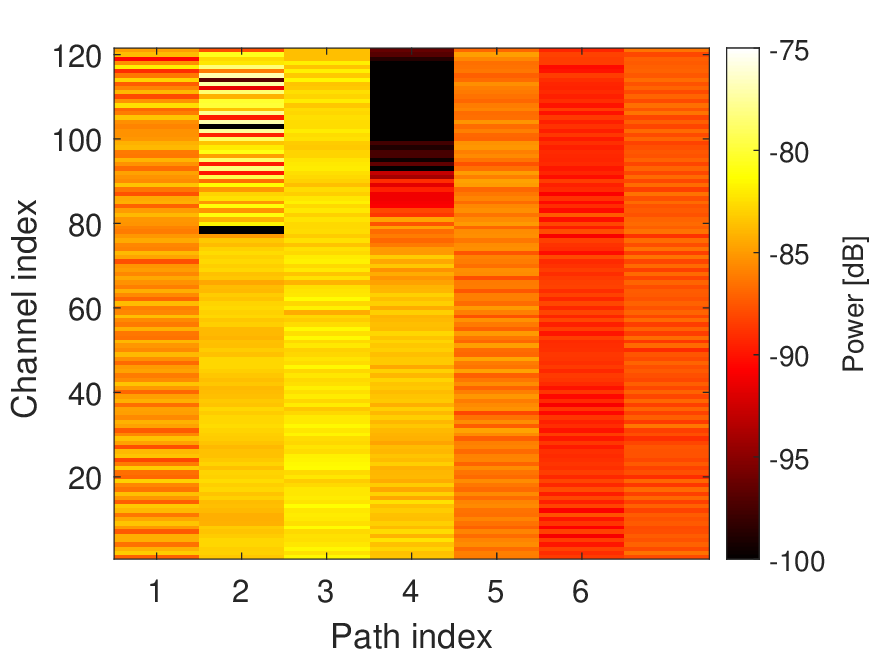}}
                        \hfill
    \vspace{-0.2cm}
        \caption{Estimated locations of scatterers, the SNS amplitude across the Rx array, and the mapping with propagation track and environment settings.}
    \label{fig:simu_result}
\end{figure*}

The problem $\mathcal{P}_1$ is a non-convex multi-object multivariate problem.
Meanwhile, the delay of multi-channels constrains the coordinates of scatterers in \eqref{eq:delay_1_1}. 
We adopt the GM-SAGE framework for coordinates search \cite{Yuan_TWC25}.
We first choose one channel as the reference channel\footnote{The reference channel can be any single-input single-output (SISO) channel of the multiple-input multiple-output (MIMO) systems.} and estimate the number of paths $L$ and the corresponding amplitudes and delays.
Then, the \textbf{E-step} and \textbf{M-step} are utilized iteratively to estimate the spatial parameters of each path.
In the $i$th iteration, for $l = 1, 2, ..., L$, 
\begin{equation} \label{eq:EM}
\begin{aligned}
\begin{cases}
 \text{E-step:}~ \hat{\mathbf{y}}_l^{(i)}  = 
    \mathbf{z}_l( \hat{\mathbf{\theta}}_{l}^{(i-1)} ) + \beta_l \left( \mathbf{y} - \sum_{l=1}^{L} \mathbf{z}_l( \hat{\mathbf{\theta}}_{l}^{(i-1)} ) \right),     \\
    \text{M-step:}~ \hat{\mathbf{\theta}}_{l}^{(i)} =  \arg \underset{ \hat{\mathbf{\theta}}_{l} }{\min} \dfrac{\left( \hat{\mathbf{y}}_l^{(i)} - \mathbf{z}_l(\mathbf{\theta}_{l} ) \right)^{H}  \left( \hat{\mathbf{y}}_l^{(i)} - \mathbf{z}_l(\mathbf{\theta}_{l} ) \right)}{\beta_{l} \sigma_0^2},
\end{cases}  
\end{aligned}
\end{equation}
where $\mathbf{\hat{y}}$ is the estimated signal of the $l$th path based on parameters of the $(i-1)$ \textbf{M-step} results.
\begin{itemize}
    \item Localization of scatterers: In the M-step, the coordinate search differs from the conventional SAGE algorithm. It utilizes the geometric constraint in \eqref{eq:delay_1_1} to directly search for the coordinates of the scatterers.
    \item Multi-bounce classification: For each path, both the one-bounce and two-bounce assumptions are operated to search the scatterers. The bouncing order is determined by the minimized value of the objective function in \eqref{eq:prob_1}. The stopping of the iteration depends on the convergence of the objective function. 
\end{itemize}
{\textbf{Blockage detection}:} Directly estimation of the sparse parameter $\mathbf{\gamma}_l$ is time-costly, e.g., using the quantities in \eqref{eq:ch_1}, the additional computational complexity is $\mathcal{O}( L 2^{MN} )$ . 
    A practical way is to estimate the equivalent amplitude as
\begin{equation}
    \begin{aligned} \label{eq:mstep_3}
      \tilde{\mathbf{\alpha}}_l^{(i)} & \triangleq  \alpha_{l}^{(i)} \mathbf{\gamma}_{l}^{(i)} \odot  \Delta \mathbf{\alpha}_{l}^{(i)} \\
     & = \left( \mathbf{z}_{l}([ \alpha_l, \Delta\mathbf{\alpha}_l, \mathbf{\gamma}_{l}, \hat{\mathbf{r}}_{l}^{(i)} ] )^H \mathbf{z}_{l}([ \alpha_l, \Delta\mathbf{\alpha}_l, \mathbf{\gamma}_{l}, \hat{\mathbf{r}}_{l}^{(i)} ] ) \right)^{-1} \\ 
    & ~~ \times \mathbf{z}_{l}([  \alpha_l, \Delta\mathbf{\alpha}_l, \mathbf{\gamma}_{l}, \hat{\mathbf{r}}_{l}^{(i)} ] )^H \hat{\mathbf{y}}_{l}^{(i)},
    \end{aligned}
\end{equation}
where in the case of partial blockage, the estimated amplitude of the estimated hidden data $\hat{\mathbf{y}}_{l}^{(i)}$ is noise. The example can be observed in Fig.~\ref{fig:simu_result}(c), where the power of the blocked channels is in the noise level.   
Therefore, the influence of SNS effects is comprehensively obtained.  

\section{Simulation and Validation}\label{vali_ray}
%
\begin{figure*}[ht]
     \centering
          \subfigure[Layout of the simulation scenario  ]{\includegraphics[width=0.325\textwidth]{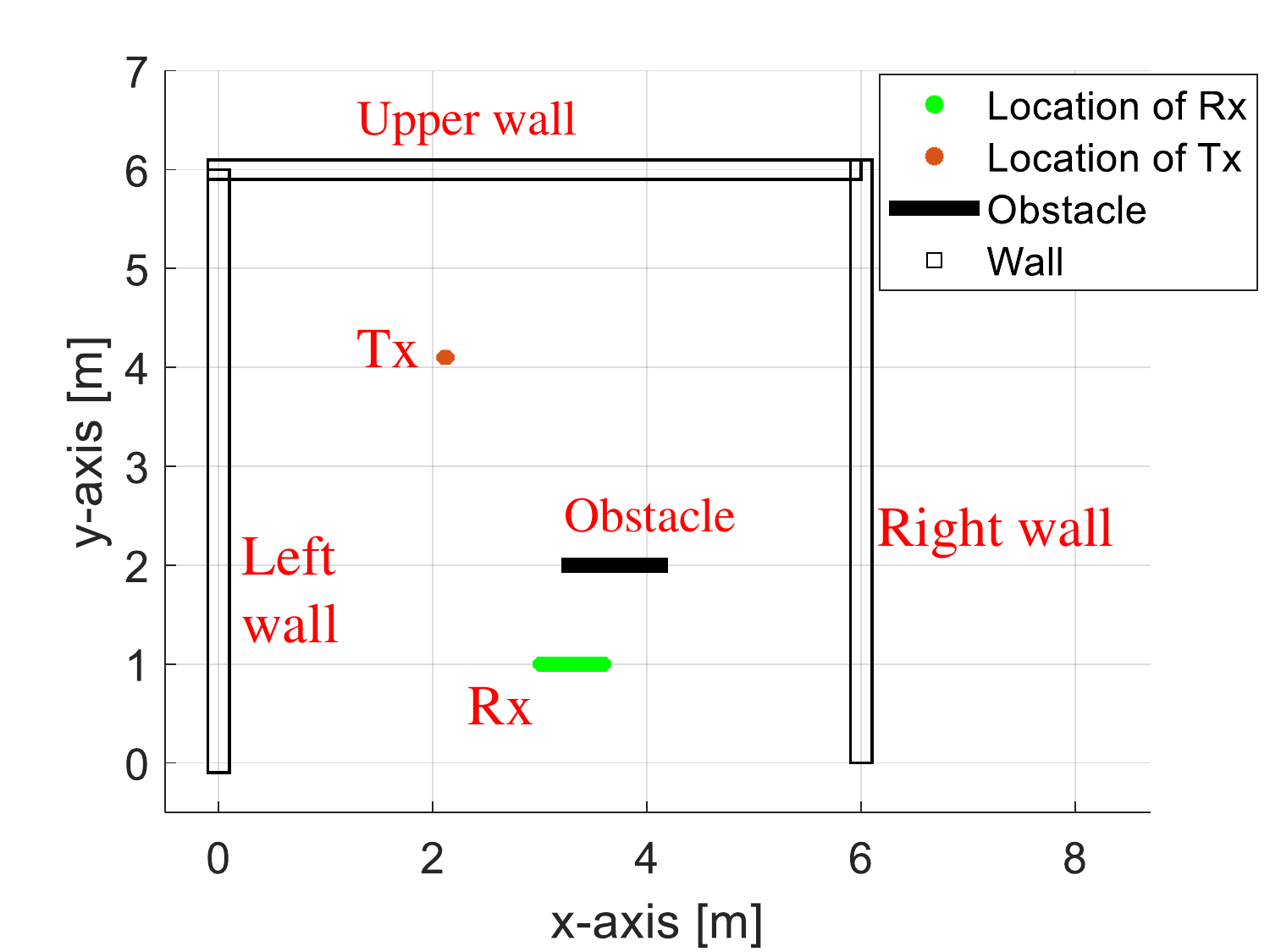}}
                        \hfill
    \subfigure[Concatenated PDP of reference Tx to all Rxs ]{\includegraphics[width=0.325\textwidth]{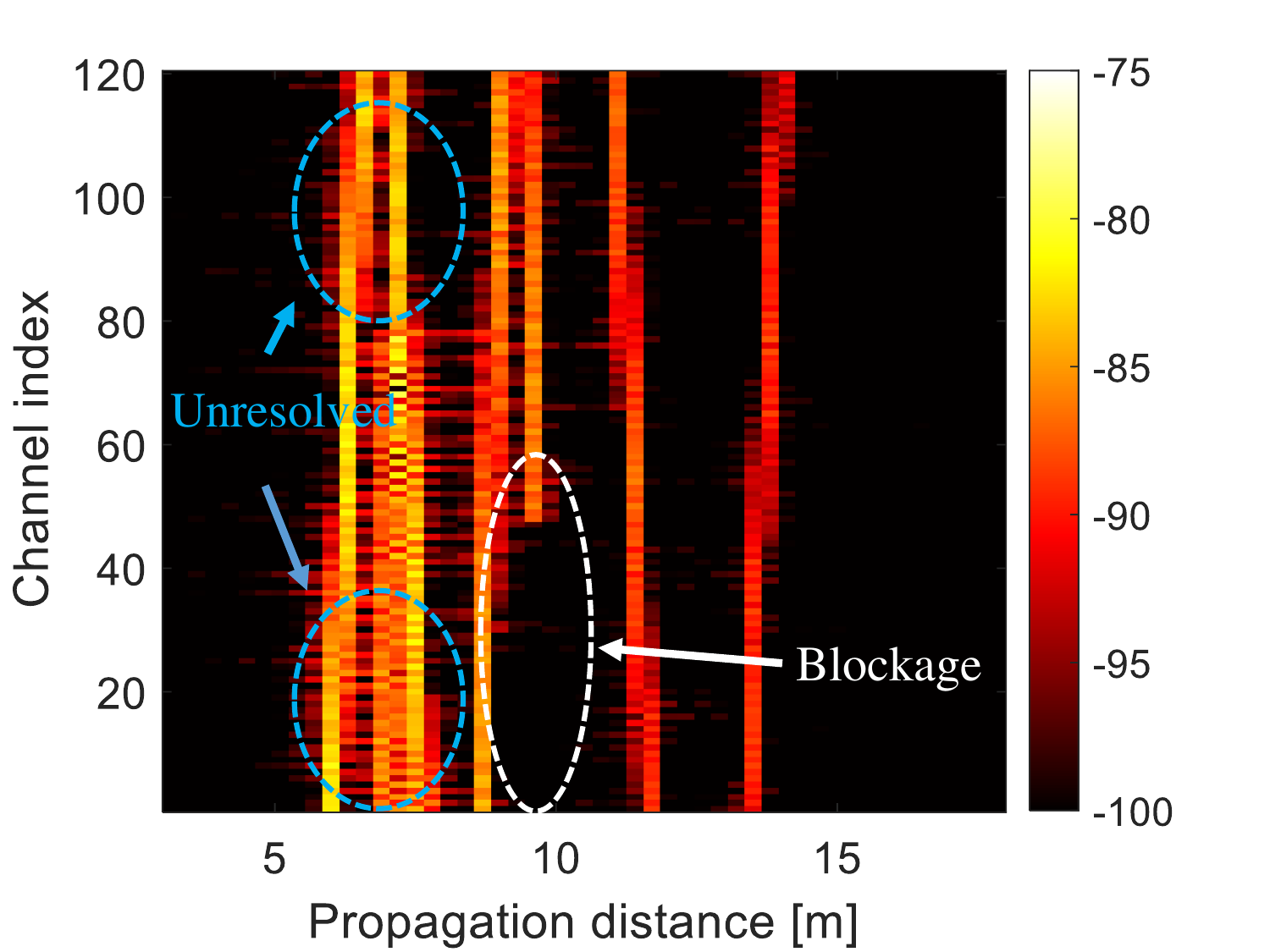}}
    \hfill
    \subfigure[Concatenated PDP of reference Tx to all Rxs ]{\includegraphics[width=0.325\textwidth]{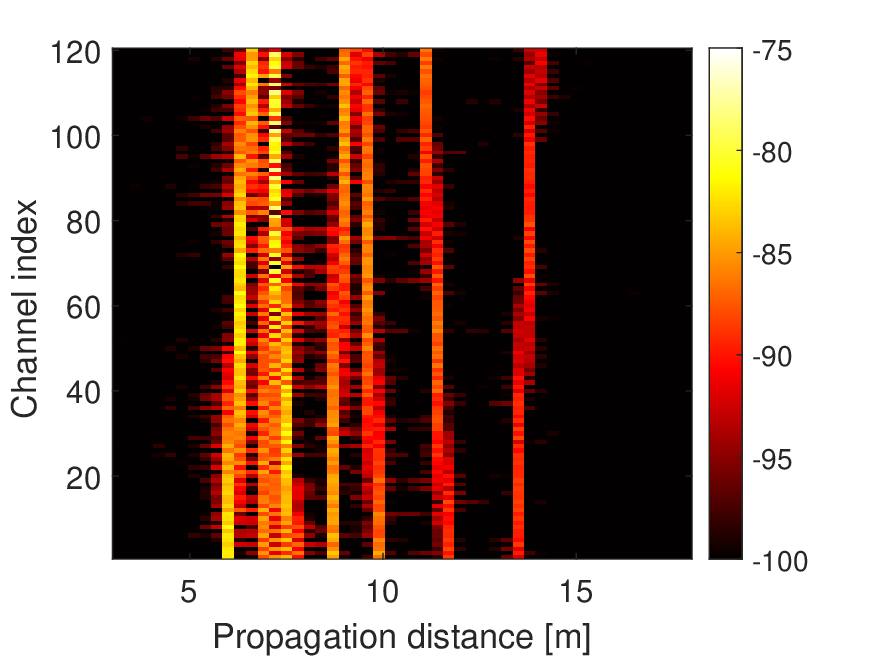}}
    \hfill
    \vspace{-0.2cm}
        \caption{Scenario and the multi-bounce channel data used in simulations.}
    \label{fig:simu_scenario}
\end{figure*}
\begin{table}[t]
    \caption{Configurations used in simulation  }
    \centering
    \begin{tabular}{lclclclc|c}
        \toprule
        Configurations &   Value  \\
        \midrule
        Central frequency $f_c$ [GHz] & $30$   \\
        Bandwidth $B$ [GHz] & $1$   \\
        Sub-bandwidth $f_s$ [MHz] & $10$   \\
        Number of sub-bands $P$ & $101$  \\
        SNR [dB] & $20$  \\  
        Grid size [m] & 0.1/0.2   \\
        Room space [m$^2$] & [${6.5 \times 6.5}$] \\
        NO. of Tx elements $\times$ Rx elements   & $16 \times 121$ MIMO \\
        Coordinates of Tx and Rx reference point & $(2.1, 4.1)$ $(3.3, 1)$ \\
        Antenna spacing & $ 0.5 \lambda $ \\
        \bottomrule
    \end{tabular}
    \label{tab:simu_configs_1}
\end{table}
\begin{figure*}[ht]
     \centering
    \subfigure[One-bounce propagation with blockage ]{\includegraphics[width=0.33\textwidth]{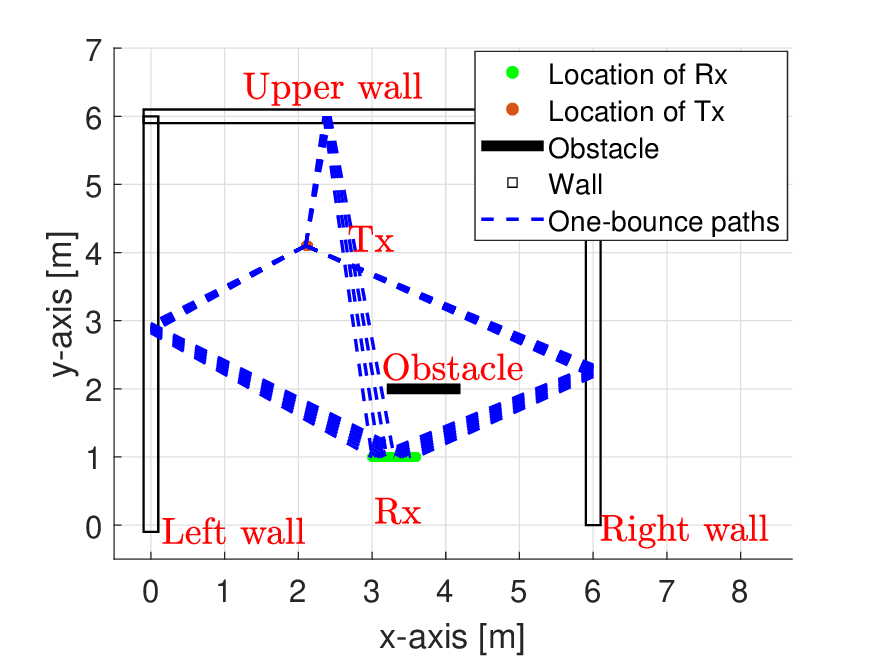}}
    \hfill
        \subfigure[One-bounce propagation without blockage ]{\includegraphics[width=0.33\textwidth]{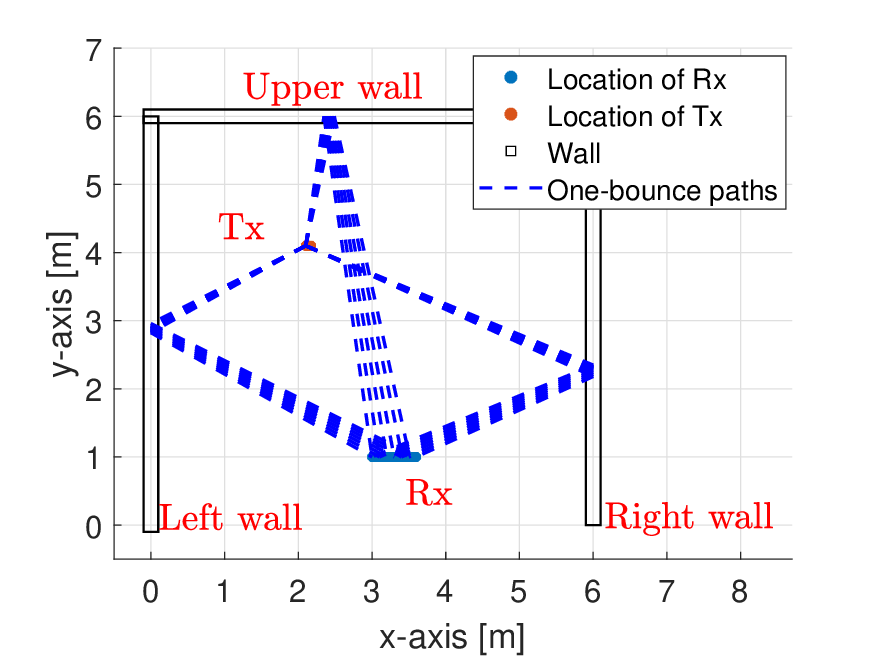}}
    \hfill
    \subfigure[The blocked track via upper wall ]{\includegraphics[width=0.32\textwidth]{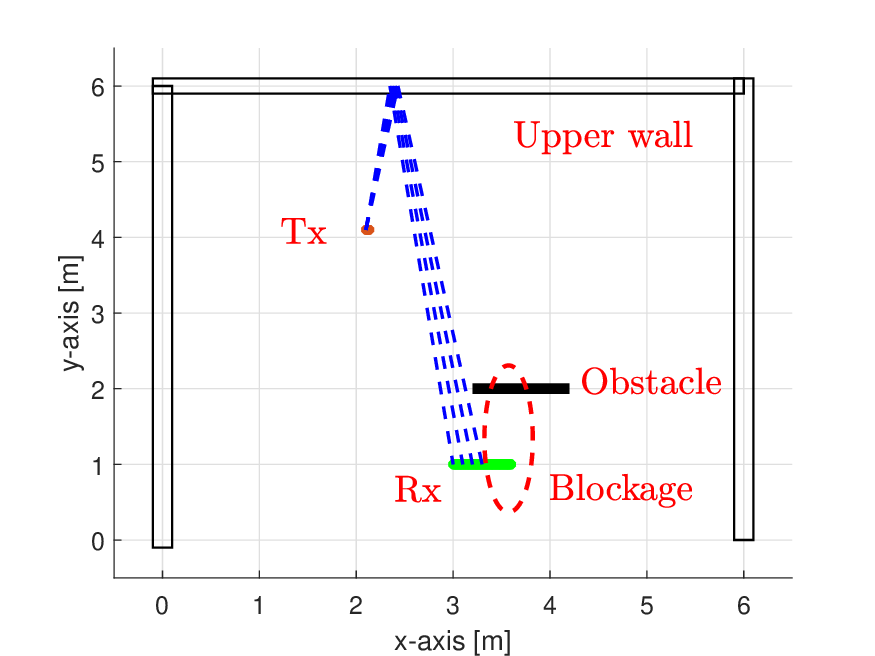}}
    \hfill
        \subfigure[Two-bounce propagation with blockage ]{\includegraphics[width=0.33\textwidth]{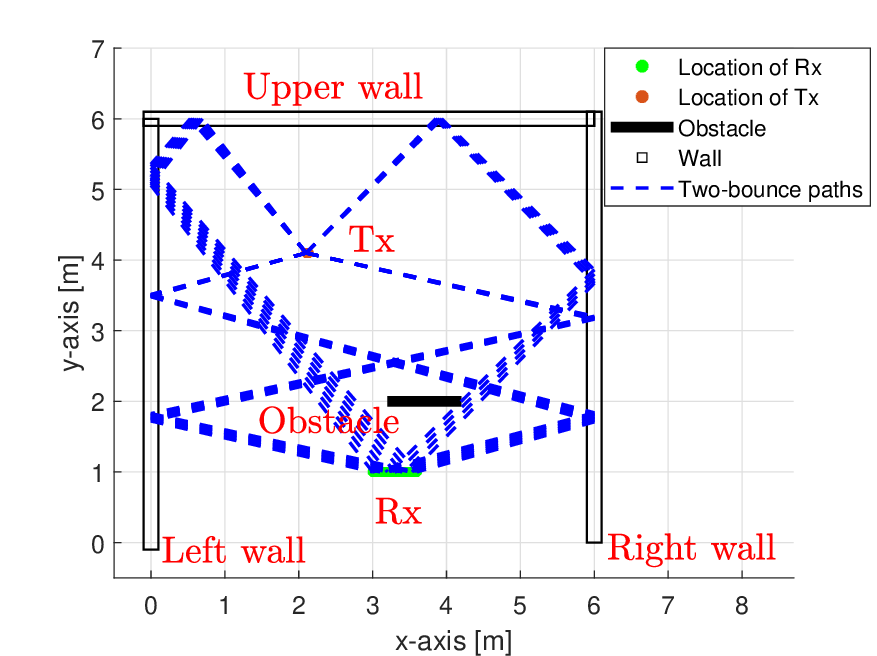}}
    \hfill
        \subfigure[Two-bounce propagation without blockage ]{\includegraphics[width=0.33\textwidth]{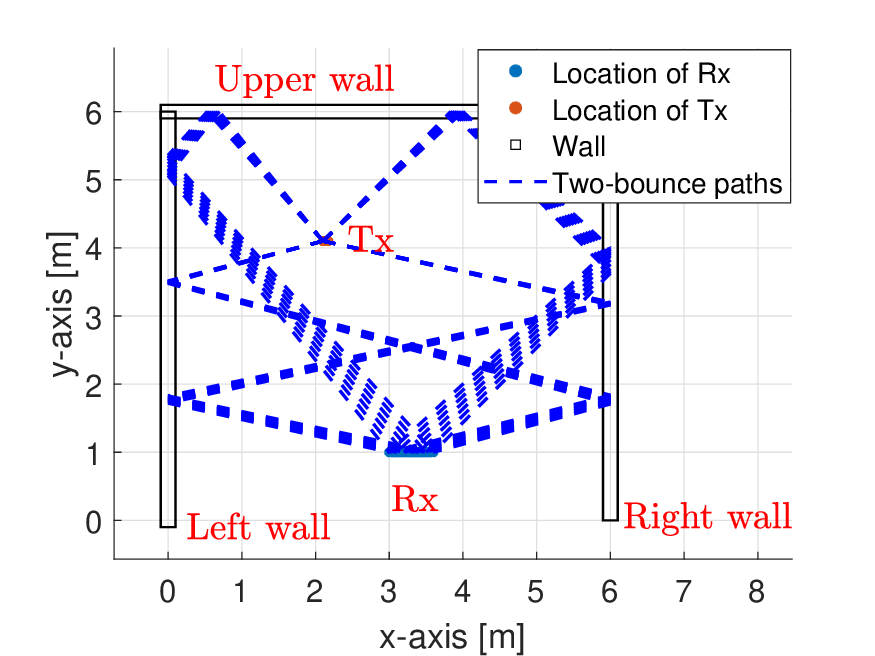}}
    \hfill
    \subfigure[The blocked track via upper and right walls ]{\includegraphics[width=0.32\textwidth]{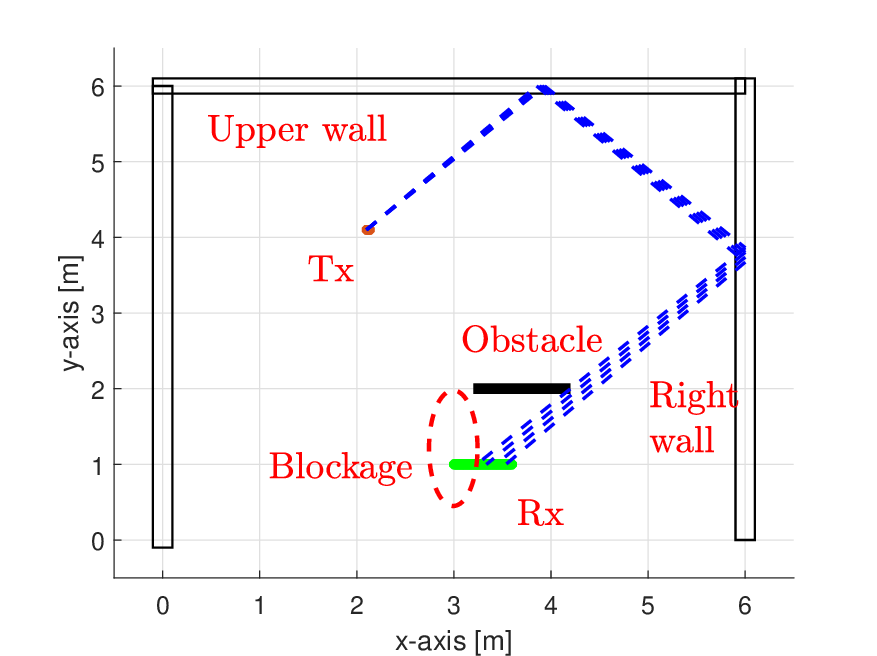}}
    \hfill
        \vspace{-0.2cm}
    \caption{Illustrations of propagation tracks of multi-bounce paths with and without blockage, where the blocked paths are highlighted in (c) and (f). }
    \label{fig:simu_propagation_track}
\end{figure*}
\begin{figure}[ht]
    \centering
\includegraphics[width=0.4\textwidth]{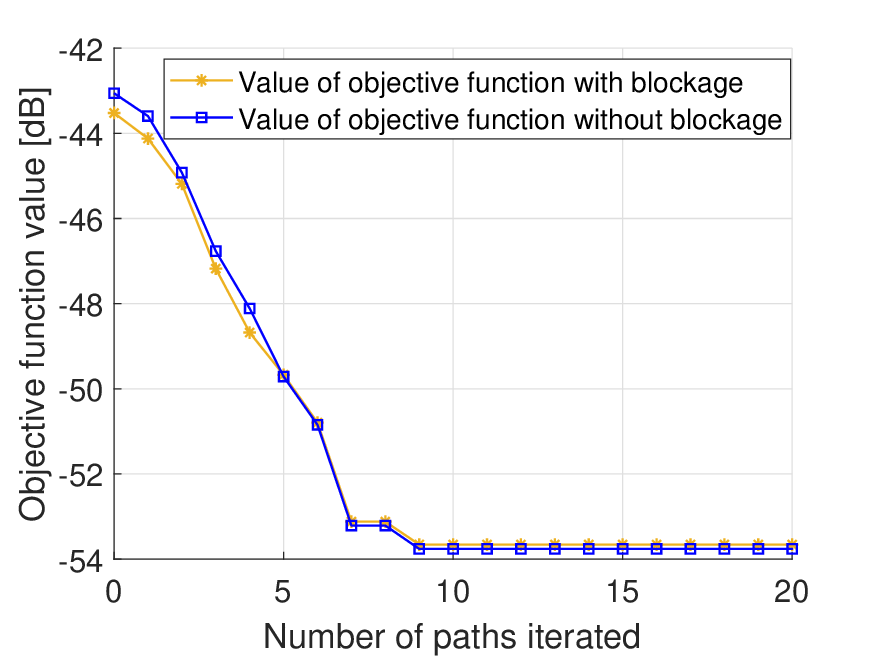}
    \caption{The convergence curve of objective function.  
    }\label{fig:simu_result_2}
    \vspace{-0.2cm}
\end{figure}
\begin{table*}[ht]
    \caption{Quantified localization error of scatterers}
    \vspace{-0.2cm}
    \centering
    \begin{tabular}{ p{1.2cm} p{0.9cm} p{1.1cm} p{1.3cm} p{1.2cm} p{2.3cm} p{2.4cm} p{2.5cm} p{2.3cm}}
        \toprule
        \midrule
        Content  & &  Left wall & \textit{Upper wall} & Right wall & Upper and left wall     & \textit{Upper and right wall} & Left and right wall  & Right and left wall \\
        \midrule
        Truth & &  $(0, 2.88)$ & $(2.46, 6)$ & $(6, 2.27)$ & $(0.64, 6)$ ~ $(0, 5.19)$ & $(3.94, 6)$ ~ $(6, 3.84)$ & $(0, 3.49)$ ~ $(6, 1.78)$ & $(6, 3.19)$ ~ $(0, 1.78)$ \\
        \midrule
        Estimation & blocked & $(0, 2.9)$ & $(2.4, 6)$ & $(6, 2.3)$ & $(0.6, 6)$ ~ $(0, 5.2)$ & $(3.8, 5.8)$ ~ $(6.2, 4)$ & $(0.4, 4.6)$ ~ $(6.2, 1.8)$ & $(6, 3.2)$ ~ $(0, 1.8)$ \\ 
        Estimation & unblocked & $(0, 2.9)$ & $(2.5, 6)$ & $(6, 2.3)$ & $(0.6, 6)$ ~ $(0, 5.2)$ & $(4, 6)$ ~ $(6, 3.8)$ & $(0.4, 4.6)$ ~ $(6.2, 1.8)$ & $(6, 3.2)$ ~ $(0, 1.8)$ \\ 
        \midrule
        Errors [m] & blocked & $0.02$ & $0.06$ & $0.20$&  $0.04$ ~ $0.01$ & $0.24$ ~  $0.26$ & $1.18$ ~  $0.1$  & $0.01$ ~ $0.02$ \\
        Errors [m] & unblocked & $0.02$ & $0.04$ & $0.20$&  $0.04$ ~ $0.01$ & $0.04$ ~  $0.04$ & $1.18$ ~  $0.1$  & $0.01$ ~ $0.02$ \\
        \midrule
        \bottomrule
    \end{tabular}
    \label{tab:error_scatter_1}
    \vspace{-0.3cm}
\end{table*}
%
\subsection{Scenario and channel data}
We validate the algorithm in a SNS channel with partial blockage scenario when employing a bistatic MIMO ELAA system of dimension $6.5 \times 6.5$~m$^2$ as shown in Fig.~\ref{fig:simu_scenario}(a). Three walls, namely left, upper, and the right wall,  and a single obstacle are included in this scenario. The key simulation parameters are shown in Table~\ref{tab:simu_configs_1}, where the  grids for searching one-bounce and two-bounce paths are $0.1$ and $0.2$~m, respectively.

We use the Image method, a commonly used algorithm in ray tracing (RT) \cite{10559769}, to generate the multi-bounce channel. 
The propagation path tracks from the reference Tx to all Rxs of one-bounce and two-bounce paths are shown in Fig.~\ref{fig:simu_propagation_track}(a) and (d), respectively. 
In particular, the one-bounce path track $\text{Tx} \rightarrow \text{upper wall} \rightarrow \text{Rx}$ and two-bounce path track $\text{Tx} \rightarrow \text{upper wall} \rightarrow \text{right wall} \rightarrow \text{Rx}$ are partially blocked; these  are shown in Fig.~\ref{fig:simu_propagation_track}(c) and (f), respectively.
As a result, the concatenated power-distance profile (PDP) of reference Tx to the $121$ receivers is shown in Fig.~\ref{fig:simu_scenario}(b), where a white dashed circle denotes the null values due to blockage. In addition, we can also observe obvious range cell migration and unresolved multipaths denoted by blue dashed circles.  
In contrast, the propagation tracks of the one-bounce and two-bounce paths without blockage are shown in Fig.~\ref{fig:simu_propagation_track}(b) and (e), respectively. The corresponding concatenated PDP is shown in Fig.~\ref{fig:simu_scenario}(c), where the power of the path with distance at $10$~m remains stationary across the entire array, because the obstacle is removed, while the unresolved multipaths are observed.  

\subsection{Estimation and environment reconstruction}
\vspace{-0.2cm}
\subsubsection{Environment reconstruction results}
Fig.~\ref{fig:simu_result}(a) and (b) show the reconstructed environment information of the one-bounce path and two-bounce path, respectively, in the partially blocked case. 
The red diamond in Fig.~\ref{fig:simu_result}(a) denotes the localized one-bounce scatterers, where the estimations match well with the propagation track;  
the red star in Fig.~\ref{fig:simu_result}(b) denotes the localized two-bounce scatterers, where most estimations match with the ground truth.
In contrast, purple diamonds in Fig.~\ref{fig:simu_result}(d) and purple stars in Fig.~\ref{fig:simu_result}(e) show the reconstructed scatterers without partial blockage.  
Table~\ref{tab:error_scatter_1} shows the difference between the estimated coordinates and the scatterers of the reference channel\footnote{As can be observed in Fig.~\ref{fig:simu_propagation_track}, the interacting point of each Tx-Rx channel has a slight difference due to the size of the receiving aperture. Hence we choose the reference channel for numerical analysis, while it is tolerated to have systematic errors. }.
\subsubsection{SNS analysis}
Using the strategy presented in \eqref{eq:mstep_3}, the SNS amplitude vector $\tilde{\mathbf{\alpha}}_l$ across the array is calculated. Fig.~\ref{fig:simu_result}(c) shows the estimated amplitude of the reference Tx and the entire Rx channels in the blocked case.
Compared with the unblocked result in Fig.~\ref{fig:simu_result}(f), the null region of path~$5$ is detected, that is, those blocked paths would not ruin the location of the scatterer. 
As a result, the two blocked paths are captured in the scatterer mapping results in Fig.~\ref{fig:simu_result}(a) and (b).
However, the blocked two-bounce path is less accurate than the unblocked one, as denoted by the black circles in Fig.~\ref{fig:simu_result}(b) and (e).   
The contrast of localization differences between the two blocked paths is in italics in Table~\ref{tab:error_scatter_1}, where the error of the one-bounce path is similar, and the two-bounce path is around $0.2$~m higher.  
Besides, Fig.~\ref{fig:simu_result_2} shows the objective function in a blocked condition can be minimized to the same level as the unblocked case. 
The above finding shows the robustness of the proposed method in scatterer reconstruction in partial blockage scenarios. 

The SNS phenomenon due to unresolved paths is also observed, as shown in the unblocked case of Fig.~\ref{fig:simu_result}(f). 
There are nulls in path~$4$, this is due to the limited resolution so that path~$3$ and $4$ in some channels are estimated as one path. This is an issue of interest worth exploring. 

\subsubsection{Outliers of two-bounce ambiguity}
We see outliers of the two-bounce path, i.e., $\text{Tx} \rightarrow \text{left wall} \rightarrow \text{right wall} \rightarrow \text{Rx}$, in both blocked and unblocked cases, shown in the orange circles of Fig.~\ref{fig:simu_result}. 
This is due to the direction ambiguity discussed in \cite{Yuan_TWC25}, which can be mitigated by beamforming at the Tx array. 

\section{Conclusion}
This paper addresses the SNS channel estimation of partially blocked ELAA systems, where the proposed GM-SAGE algorithm is used to locate scatterers of multi-bounce propagation and reconstruct the environment.
Specifically, the SNS amplitude of each channel is obtained by minimizing the chosen objective function; hence the blocked channels do not degrade the estimation of scatterer locations and the validation cases show the robustness of the algorithm.  Further, we also observed SNS due to unresolved multipaths in the delay domain; this is left for future investigation. 
\bibliographystyle{IEEEtran}
\bibliography{ref.bib}
\end{document}